# DIAMOND Control System Outline Design

M. T. Heron, A. J. Duggan, B. G. Martlew, A. Oates, P. H. Owens, V. M. Smithers, P. D. Woods
CLRC Daresbury Laboratory, Warrington, Cheshire, WA4 4AD, United Kingdom

*Abstract*

DIAMOND is a new synchrotron light source currently being designed in the UK. The control system for DIAMOND will be a site wide monitoring and control system for the accelerators, beamlines and conventional facilities. Initial work on the control system has selected EPICS as the basis for the control system design.

The requirements of the control system are presented. The technical solutions being considered to interface EPICS to the equipment being controlled are discussed together with the requirements for configuration and management of a large EPICS installation. Options being considered for the procurement, installation and commissioning of the control system are also presented.

## 1   INTRODUCTION

DIAMOND is a 3rd generation, 3GeV synchrotron light source currently being designed in the UK. The storage ring is based on a 24-cell double bend achromatic lattice of about 560m circumference. It uses a full energy booster synchrotron and linac for injection. The spectral output is optimised for high brightness up to 20keV from undulators and high flux up to 100keV from multipole wigglers. Initial construction includes seven photon beamlines.

The project is currently in a design and specification phase, with build and procurement starting in April 2002 and operation scheduled from Sept 2006. Details of the project status are described in [1].

## 2   CHOICE OF CONTROL SYSTEM TOOL KIT

Four options for the control system tool kit were reviewed during 1999: TACO/TANGO from the ESRF [2,3], commercial control systems, a development of the SRS control system[4] and EPICS[5]. The conclusion of this was that any one could deliver a working control system[6]. Following on from this a detailed analysis of control system tool kit requirements showed that EPICS offered advantages; in terms of its application to other accelerator projects, a larger base of developed drivers and performance/functionality.

## 3   CONTROL SYSTEM SCOPE

The defined scope for the control system is: that it will be a site wide monitoring and control system for the accelerators, beamlines and conventional facilities. The control system will extend from the interface of the equipment being controlled through to the operator. It will include all hardware and software between these bounds: computer systems, networking, hardware interfaces and programmable logic controllers.

The control system will not include any control or data acquisition for the experimental stations. It further will not include the personnel protection system, which is a separate system that will be monitored by the control system.

### 3.1 Operational Requirements

The control system will have to support a facility operating in excess of 5000 hours per year but with high control system availability required at all times. The facility is expected to operate for around twenty years.

The operational functions to be provided by the control system must include the routine operation of the facility by the operator, support for the technical groups to diagnose equipment and for the physics group to carry out experiments in characterising the facility.

### 3.2 Interface Requirements

Considering typical systems across each of the accelerators and the 7 beamlines there will be around 4000 physical devices that will require interfacing into the control system. The estimated numbers of interface types, not necessarily channels, for these devices is shown in table 1.

Table 1: Numbers of interface types

| Interface type | Quantity |
|---|---|
| Analogue | 1677 |
| Digital | 1111 |
| VME back plane | 73 |
| Video | 8 |
| GPIB | 2 |
| Serial | 2116 |
| System Integration | 8 |

# 4 DIAMOND CONTROL SYSTEM

The DIAMOND control system will adopt the standard two-layer architecture with workstations as the client and VME crates as the server. There will be no field bus to a third layer, but extensive use will be made of a field bus or serial interfaces from layer two, to the equipment.

## 4.1 EPICS Tool Kit

The current development work is using EPICS 3.13.4 and VxWorks 5.4, which is providing a stable environment. The versions of EPICS and VxWorks to be used for the installed system will be fixed later in the design phase.

## 4.2 Equipment Interface

The interface from the control system to the equipment will be through VME64X Input Output Controllers (IOCs). IOCs will be installed for each technical system e.g. Vacuum and for each geographical area i.e. per cell on the storage ring, giving a total of around 212 IOCs, as shown in table 2. These will use PPC processor boards, Industry Pack carriers and modules as the principal interfaces. The preferred interfaces to equipment will be analogue, digital and serial (RS232, RS422 etc). Support will also be provided for at least one field bus to interface PLCs. The choice of field bus is currently undecided but preference will be for an open standard field bus. Each IOC will also contain an event receiver for synchronous operation and accurate time stamping of data. This combination allows for a high density of IO from a compact VME system, with most requirements being satisfied by four VME modules housed in seven slot crates.

Table 2: Numbers of IOCs

| IOCs | Linac | Tx Paths | Booster | SR | BLs (7) |
|---|---|---|---|---|---|
| Main Magnets | | 1 | 1 | 25 | |
| Steering Magnets | | 1 | 1 | 24 | |
| RF | 1 | | 2 | 5 | |
| Vacuum | 1 | 2 | 4 | 48 | 7 |
| Diagnostics | 1 | 2 | 4 | 25 | |
| Pulsed PSUs | | | 1 | 1 | |
| Personnel Safety | 1 | | | 2 | 2 |
| Vessel protection | | | | 24 | |
| Rad. Monitors | 1 | | | 2 | 2 |
| IDs or Motors | | | | 7 | 7 |
| Misc | | | | 5 | 2 |
| TOTALS | 5 | 6 | 13 | 168 | 20 |

## 4.3 Serial Interfaces

The decision has been taken to support the interfacing of equipment through serial interfaces. This offers benefits in being able to use commercial off the self equipment, reduced wiring, better correlation with local equipment values, and reduced installation and commissioning times. It is recognised that serial data processing will impose an IOC CPU overhead and supporting multiple vendor specific protocols will impose a development overhead.

## 4.4 Equipment protection

Equipment protection is included within the control system where required, with the control system issuing hardwired "permit to operate" signals to equipment. Three levels of protection, determined by assessment of damage/cost caused by failure, are: High Integrity, provided in hardware; Routine Interlocking, provided by PLCs; Prudent Operational Limits, provided by IOCs.

Global interlocks are required to protect the storage ring vessel and dipole magnets. These will be generated at each control and instrumentation area, converted and sent via the control system computer room over the fibre optic cables to the RF plant and dipole PSUs respectively.

## 4.5 Consoles

The consoles will be workstations and for cost reasons these are likely to be PCs running Linux. There is also a requirement to provide a control interface to Win32 based applications.

## 4.6 Central Servers

Central servers will be provided for development, applications, relational database, intranet and internet, archiver, alarm management and machine model and IOC booting. At each control and instrumentation area there will be local a boot server for the IOCs. These will be the primary IOC boot servers to minimise network load from IOCs rebooting. Consideration is being given to standardising all servers on Linux.

## 4.7 Applications

The principal application requirements can be met through the standard EPICS tools, control panels through MEDM, alarm management through Alarm Handler, archiving through Channel Archiver together with scripting languages for rapid application development.

A standard application for viewing and controlling channels in a tabular form is planned and an application for applying settings to a large number of the channels, to sequence the accelerator from one mode to another will be developed.

Applications will interface to the control system through the CDEV[7] abstraction layer. This will facilitate the integration of dynamic control data from Channel Access and persistent data stored in a relational database (RDB).

Soft records are currently being used to simulate the channels available in the final control system and so facilitate early application development.

### *4.8 Configuration and management*

Capfast[8] is currently being used for IOC DB configuration but consideration is being given as to whether Visual DCT[9] offers a greater long-term future.

The RDB is planned to manage conventional control system information together with the control system configuration. All IOC DBs will be generated out of the RDB so ensuring that there is a central repository for all control parameters. Work required to link the EPICS database build process together with RDB is current being defined.

### *4.9 Physical structure*

The control system will interface to the other technical systems at 48 control and instrumentation areas covering the linac, booster, transfer lines, plant rooms, storage ring and beamlines. For the storage ring there will be one control and instrumentation area per cell, thereby ensuring each cell is self-contained.

### *4.10 Network*

A network infrastructure will connect the control system computer room to each of the 48 control and instrumentation areas with single and multi mode fibres. The fibre will provide two computer networks: one for control system and one for other computer systems to enable effective management of traffic on the control system network. Each network will use a central switch in the control system computer room and the control network will have secondary switches at each control and instrumentation areas. The fibres will further be used for event distribution, global interlocking and beam position feedback system

## 5 CONSTRUCTION

The detailed programme of work for construction phase of the project is currently being planned with latest options presented.

### *5.1 Procurement*

The sub system below the IOC, principally PLC based, will be procured as design and build contracts from conventional systems houses. The procurement of complete technical sub systems, e.g. Linac with EPICS IOCs fully integrated, is being discussed with likely vendors.

The majority of IOCs will be procured on a component basis and integrated in house. The procurement of VME modules and instrumentation with EPICS device support is being investigated, with one VME manufacturer already developing EPICS support to our requirements.

### *5.2 Commissioning.*

Commissioning is planned to progress through the linac, booster and storage ring. The storage ring cells are based on three girders, which support the vessel and magnets. These will be constructed offline and at this stage tested in conjunction with the control system. By ensuring each cell of the storage ring is self-contained, as the technical systems are installed the control system for that cell can be set to work and commissioned.

## 6 CONCLUSION

The structure for the DIAMOND control system is now being defined. With the control system well specified early on in the project the detail design can be resolved to ensure a high level of functionality available for day one commissioning.